\documentclass[iop]{emulateapj}     % two column  

\usepackage{txfonts}
\usepackage{graphicx}

\usepackage{graphicx}
\usepackage{soul}

\usepackage{natbib}

\usepackage{natbib}

\begin{document}

 \title{A hot downflowing model atmosphere for umbral flashes and the physical properties of their dark fibrils.} 

   \author{V. M. J.  Henriques$^{1,2}$, M.  Mathioudakis$^{1}$, H. Socas-Navarro$^{3}$,  J. de la Cruz Rodr\'iguez$^{4}$}
\affil{$^1$ Astrophysics Research Centre, School of Mathematics and Physics, Queen's University Belfast, BT7 1NN, Northern Ireland, UK \email{v.henriques@qub.ac.uk}}
\affil{$^2$ Institute of Theoretical Astrophysics, University of Oslo, PO Box 1029 Blindern, 0315 Oslo, Norway}
\affil{$^3$ Instituto de Astrof\'isica de Canarias, Avda v\'ia L\'actea S/N, 38205 La Laguna, Tenerife, Spain}
\affil{$^4$ Institute for Solar Physics, Department of Astronomy, Stockholm University, AlbaNova University Centre, 106 91 Stockholm, Sweden}

\date{Received 2017 March ; Accepted 2017 June}                          
  
  \begin{abstract} We perform NLTE inversions in a large set of umbral flashes, including the dark fibrils visible within them, and in the quiescent umbra by using the inversion code NICOLE on a set of full Stokes high-resolution Ca~II~8542~\AA\  observations of a sunspot at disk center. We find that the dark structures have Stokes profiles that are distinct from those of the quiescent and flashed regions. They are best reproduced by atmospheres that are more similar to the flashed atmosphere in terms of velocities, even if with reduced amplitudes. We also find two sets of solutions that finely fit the flashed profiles: a set that is upflowing, featuring a transition region that is deeper than in the quiescent case and preceded by a slight dip in temperature, and a second solution with a hotter atmosphere in the chromosphere but featuring downflows close to the speed of sound at such heights. Such downflows may be related, or even dependent, on the presence of coronal loops, rooted in the umbra of sunspots, as is the case in the region analyzed. Similar loops have been recently observed to have supersonic downflows in the transition region and are consistent with the earlier "sunspot plumes"  which were invariably found to display strong downflows in sunspots. Finally we find, on average, a magnetic field reduction in the flashed areas, suggesting that the shock pressure is moving field lines in the upper layers. 

\end{abstract}

   \maketitle

\section{Introduction}

Present day inversion codes like SIR \citep{1992ApJ...398..375R}, SPINOR \citep{2000A&A...358.1109F,2012A&A...548A...5V}, STIC \citep{2016ApJ...830L..30D}, and NICOLE \citep{2015A&A...577A...7S} allow us to compute semi-empirical model atmospheres in the millions per dataset, covering a wide-range of dynamically evolving structures. High resolution data allow the study of small-scale variations, in space and time, of highly dynamic fine structured events such as umbral flashes, as well as providing new insights into earlier modeling work where dynamic phenomena tend to be averaged together. 

Umbral flashes were first characterized by \cite{1969SoPh....7..351B} in the Ca~II~H and K lines, who proposed that these are acoustic shocks based on their propagation across field lines with such propagation confirmed that same year by \cite{1969SoPh....7..366W}. They display a shock-characteristic saw-tooth pattern, known as "z-pattern" in early literature \citep{1984A&A...135..188T}. These seem to steepen directly from the photospheric 3-minute oscillations \citep{1985AuJPh..38..811T,2006ApJ...640.1153C}. With an atmospheric model that included a forced piston, \cite{2010ApJ...722..888B} apply 1D NLTE radiative transfer using MULTI \citep{MatsUppsala1986}, to successfully reproduce the flash intensity profiles in Ca~II~H as well as the saw-tooth pattern. The latter can have peak to peak  velocities of up to 15~km~$s^{-1}$  \citep{2003A&A...403..277R,2014ApJ...786..137T}, as observed also in Ca~II~H. This pattern is also observed in He I, with peak to peak amplitudes of up to 11~km~$s^{-1}$ \citep{1986ApJ...301.1005L}. Further up, umbral flashes seem to be precursors to the running waves observed in the transition region \citep{2015ApJ...800..129M,2016PhDT........15L}. Latency between different chromospheric lines indicates upward propagating waves and energy but the estimates of the mechanical energy do not seem sufficient to compensate for the radiative losses \citep{1981A&A...102..147K}. For a recent review of the properties of umbral flashes in a broader context (such as their connection to running penumbral waves) see \cite{2015SSRv..190..103J} and \cite{2015LRSP...12....6K}. 

Early umbra semi-empirical models featuring a chromosphere include \cite{1986ApJ...306..284M} and more recent automated inversion work such as \cite{2013A&A...549A..24B} and \cite{1997ASPC..118..202W}, who used the SIR code \citep{1992ApJ...398..375R} in the Ca~II~H line and the 6302 line pair respectively. Both include magnetic field values for the umbra of sunspots above 2000 G in the chromosphere indicating little change in the magnetic flux from the photosphere to the chromosphere. This agrees with the observations of \cite{1997ASPC..118..202W} who find only a small reduction of the magnetic field strength, 2400 to 2100 G, at $\log \tau_{500} = -2.8$, for an umbral average of the inverted data. For a more comprehensive review of earlier umbral models please refer to \cite{2003A&ARv..11..153S}. However, the chromosphere in these models was either extrapolated or directly inferred from LTE modeling.  

The first spatially continuous 3D model of a sunspot, produced using NLTE inversions in a chromospheric line, was that of \cite{2013A&A...556A.115D} who found enhancements of up to 1000~K in the flashed atmosphere and a magnetic field oscillation, modulated by the flashes, using the weak-field approximation. Inversions in the chromosphere tend to yield highly inhomogeneous results, both vertically and spatially \citep{2007ApJ...670..885P} which presents challenges to interpretation. For a recent review on the state of the art of inversions see \cite{2016SSRv..tmp...73D}.

\cite{2000ApJ...544.1141S} found abnormal Stokes profiles in the flashed phase itself that hinted at unresolved features. \cite{2000Sci...288.1398S,2001ApJ...550.1102S} successfully fitted (in NLTE) flashed and quiescent profiles with a linear combination of a two-component (unresolved) atmosphere, one upflowing and the other gently downflowing, where the proportion of each component would vary in time to reproduce the observed profile variation. In the infrared, \cite{2005ApJ...635..670C} studied time-series of flashes and found evidence for two components both in a temporal and spatial sense, which was interpreted as evidence for separate channels for upflows and downflows. \cite{2007PASJ...59S.631N} found a node-like feature over the dark umbra where power was suppressed at 5.5~mHz powermaps as seen in Ca~II~H. Also in Ca~II~H, \cite{2009ApJ...696.1683S} found dark fibrils in high-resolution Hinode filtergrams, with longitudinal horizontal projections as long as 2000~km. This observation was confirmed in a different sunspot by \cite{2013A&A...557A...5H}. \cite{2015A&A...574A.131H} found that such streaks could be even longer and stable for at least two flashes, with evidence for the presence of the same fibrils over more than three flashes. Two populations of features seemed to exist, with the longest observed feature extending over the penumbra and showing signs of a change of inclination when crossing the umbra-penumbra boundary, and the smaller fibrils having a partial match with H-alpha features, with properties consistent with short-dynamic fibrils. The latter is in agreement, at least partially, with the work where short dynamic fibrils were discovered \citep{2013ApJ...776...56R} and in agreement with the likely identical H-alpha spikes \citep{0004-637X-787-1-58}. Why such dark fibrils, as well as the likely related short-dynamic fibrils, are visible at all is still an open question due to an absence of an identified source of inhomogeneities in the chromosphere of the umbra of sunspots. Other commonly occurring umbral fine structure, umbral microjets, have been observed in emission in Ca~II~H by \cite{2013A&A...552L...1B}, and likely have a non-direct relation with short dynamic fibrils \citep{2017arXiv170505617N}.

In this work we have produced and analyzed an array of umbral models that reproduce the observed time-series of Stokes profiles. Limited attention is given to any one single fit and we make use of the large amount of atmospheres, generated by the inversion procedure, to gain insight into the above phenomena and the possible source of inhomogeneities in the umbra.   

%%%%%%%%%%%%%%%%%%%%%%%%%%%%%%%%%%%%%%%%%%%%%%%%%%%%%%%%%%%%%%%%
\section{Observations and Data Processing}
\label{sect:setup}

We used the CRisp Imaging SpectroPolarimeter \citep[CRISP;][]{2006A&A...447.1111S,2008ApJ...689L..69S} instrument,
at the Swedish 1-m Solar Telescope \citep[SST;][]{2003SPIE.4853..341S}. Adaptive optics were used, including an 85-electrode deformable mirror which is an upgrade of the system described in \cite{2003SPIE.4853..370S}. 
All data were reconstructed with Multi-Object Multi-Frame Blind Deconvolution \citep[MOMFBD;][]{2002SPIE.4792..146L,2005SoPh..228..191V}, using 82~Karhunen-Lo\`{e}ve modes sorted by order of atmospheric significance and $88\times88$ pixel subfields. 

A prototype of the data reduction pipeline published by \cite{2015A&A...573A..40D} was used before and after MOMFBD. This includes the method described by \cite{2012A&A...548A.114H} for consistency between the different LC states and wavelengths, with destretching performed as in \cite{1994ApJ...430..413S}. 

\begin{figure}[!htb]
\begin{center}
\includegraphics[width=7.1cm]{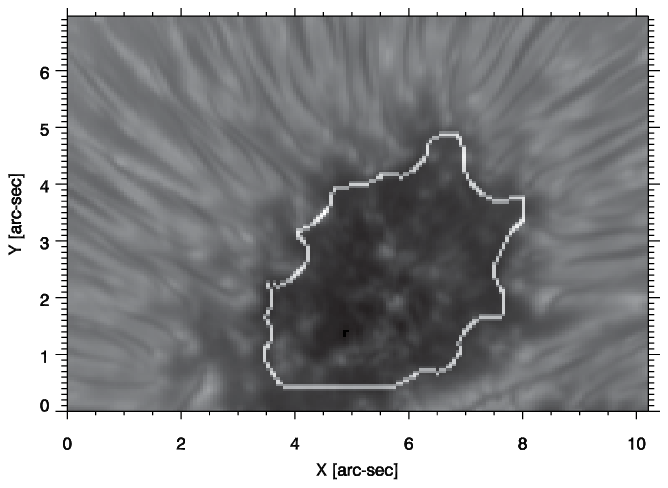}
\includegraphics[width=7.1cm]{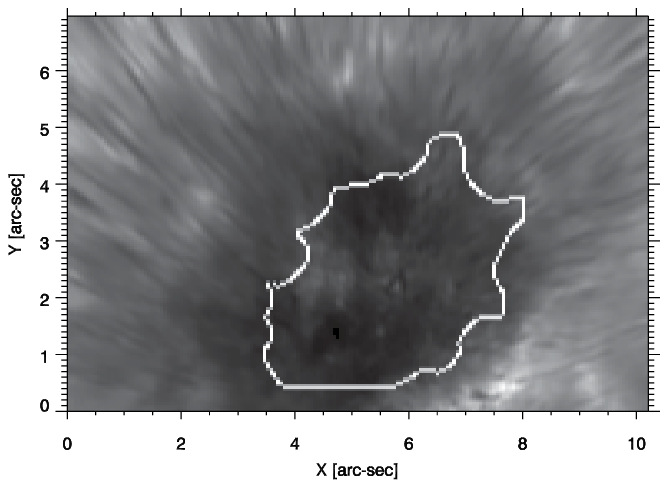} 
\includegraphics[width=7.1cm]{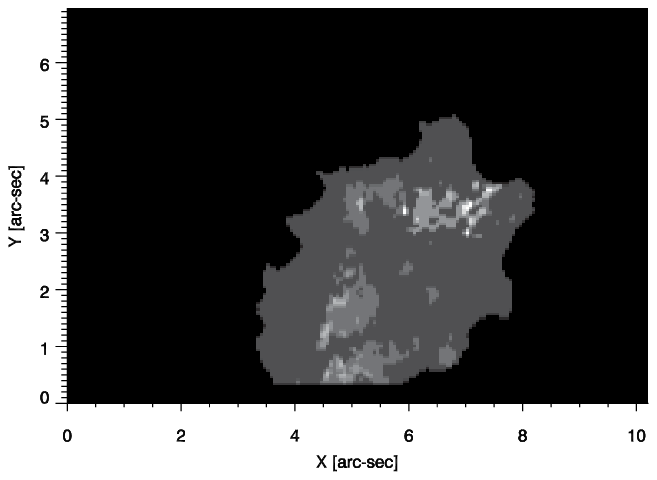}
\end{center}
\caption{Top panel: context image in the wing of the Ca~II~8542~\AA\ line at +942~m\AA\ from line core. Second panel: context image at -73~m~\AA\ from line core during a flash. Two flashed regions and dark fibrils are visible. The contours indicate the border of the analysed area (umbra).  Third panel: the grey contour indicates the umbra and brighter grey tones indicate the locations of the flashed pixels detected. Each level of brighter grey indicates that the pixel was detected as a flashed pixel for an additional scan in the time series.}
\label{context}
\end{figure}

The observations were normalized to the intensity of the continuum levels by fitting the FTS atlas profile \citep{1999SoPh..184..421N}, convolved with the CRISP wavelength profile, to an average of the quiet-Sun profile computed from multiple scans (averaging over every fifth scan).  The main target of the observations and of the inversions was the largest umbra of the main sunspot in the NOAA 12121 active region, when it was close to disk center (x=76\arcsec,y=46\arcsec), between 10:43 and 11:23~UT on the 28th of July 2014. Figure~\ref{context} shows the inverted field-of-view (FOV). 

\begin{figure}[!htb]
\begin{center}
\includegraphics{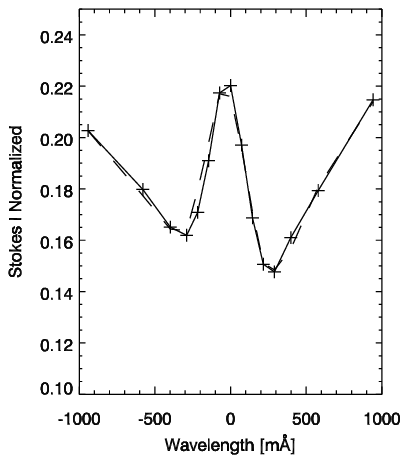} 
\includegraphics{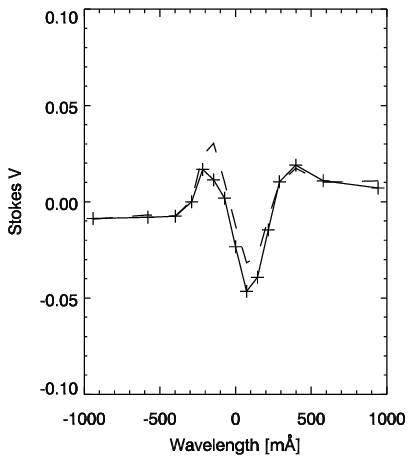}  
\end{center}
\caption{Observed (dashed) and synthetized profiles (solid) from a flashed pixel that were considered to be on the limit of what constitutes an acceptable fit. Any result with a worse goodness of fit was discarded. The crosses indicate the observed wavelengths.}
\label{limit_ths}
\end{figure}

The spatial sampling is 0$''$.0592 per pixel, with the spatial resolution reaching up to 0$''$.18 over the FOV of 41$\times$41~Mm. The Ca~II~8542~\AA\ line was sampled from -290~m\AA\ to +290~m\AA\ in steps of 73~m\AA\ (as measured from the averaged observed core of the line in a quiet area at disk center) and at  -942m\AA, -580m\AA, -398m\AA, +398m\AA, +580m\AA, and +942 m\AA, for a total of 15 wavelengths. The observed line positions can be seen as crosses in Figure~\ref{limit_ths}. Full Stokes polarimetry was achieved by using four liquid crystal (LC) states and a demodulation scheme that included a calibration of the optics on table from the telescope primary focus to the science focus, taken less than 3 hours from the observations, and a telescope model spanning the primary focus up to and including the primary lens. The latter was produced from calibrations taken the same year and includes daily variations of the telescope modulation (see  \cite{2015A&A...573A..40D} and \cite{2011A&A...534A..45S}). Our inversions overestimated the magnetic field at all heights and all pixels by roughly a constant factor due to an underestimated direct transmission in Stokes~V in the telescope model which was discovered only close to the submission of this work. Thus all values were compensated, post-inversions, with the ratio of the used transmission and that of the correct transmission (0.605) by assuming the weak field approximation (i.e. that the magnetic field strength is proportional to the amplitude of the Stokes V profile). While the absolute value of the magnetic field seems to be within those measured in previous literature (i.e. between 2~kG and 3~kG in the umbra) we abstain from making claims about the absolute values of the magnetic field, but do study relative differences. Two scans were reconstructed together using MOMFBD for a total of 28 seconds per scan and 14 frames per wavelength position per LC state. This binning in time at the level of the reconstruction was done for two reasons:  in order to increase signal to noise in the Stokes profiles and so that periods of poor seeing, of the order of a second and thus potentially longer than the acquisition time of a single wavelength, would have a much reduced probability of affecting the final line profile in areas of high spatial gradient (such as in dark fibrils). Combining two scans at the reconstruction level benefits from the fact that MOMFBD overweights images, for the same wavelength, where the quality is highest. Further, the observations were binned spatially in a 2x2 fashion to increase signal and reduce noise. This lead to a pixel size of 0.$''$12 which is just below the maximum resolution of the SST at this wavelength (0.$''$18) even if this means that we forego Nyquist sampling of the resolution element.

\begin{figure}[!htb]
\begin{center}
\includegraphics[width=7.1cm]{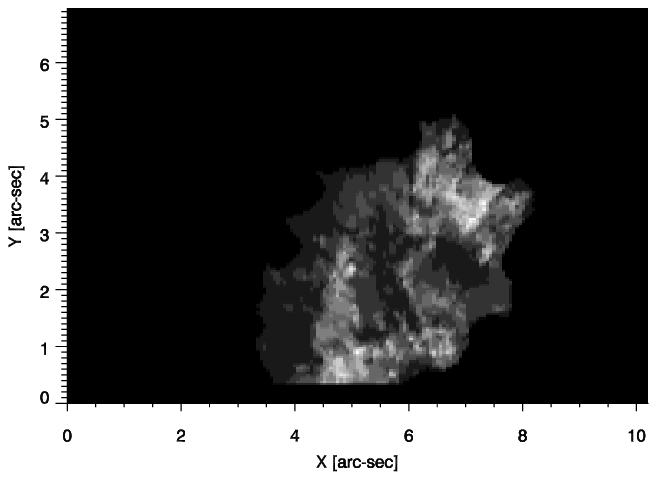}
\includegraphics[width=7.1cm]{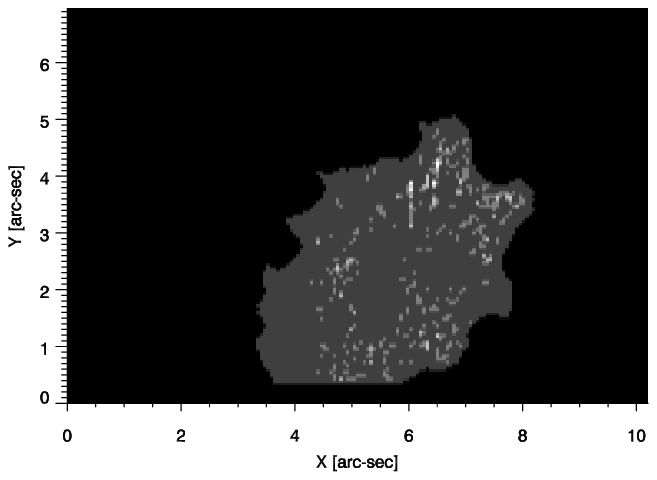}
\end{center}
\caption{Top panel: the grey contour indicates the umbra and brighter grey tones indicate the locations of the flashed pixels detected using a relaxed intensity threshold below what was used for analysis and not including any quality-of-fit thresholding. Each level of brighter grey indicates that the pixel was detected as a flashed pixel for an additional scan in the time series. Bottom panel: same grey scale scheme but indicating detected dark fibrils as described in the text. Note that the regions are much smaller, elongated, and form counterparts to the holes in the map of the top panel. The dark fibril map shown is already selected for fit quality and thus these are the pixels used for the density plots in Figure~\ref{ucf} and Figure~\ref{bdens}.}
\label{context2}
\end{figure}

The inversion code NICOLE \citep{2015A&A...577A...7S} was used in NLTE mode with a three angle Gaussian quadrature for the radiation field. A 5-level plus continuum Ca~II atom was used as in \cite{2013A&A...556A.115D}. The Wittman approach to the equation of state was selected to compute the unfitted thermodynamical variables \citep{1974SoPh...35...11W}. The cubic delo-Bezier solver was selected for the radiative transfer \citep{2013ApJ...764...33D}. We include the effect of Ca II isotopic splitting in the inversions \citep{2014ApJ...784L..17L}. 

A scheme of multiple inversion cycles with ever increasing nodes was used as suggested by \cite{1992ApJ...398..375R}. The atmospheres were smoothed in between each cycle, both horizontally and vertically, but the smoothing itself was performed only on the perturbations from the previous cycle. For example, if NICOLE found no way to improve an atmosphere in any given cycle, then no smoothing was applied and the initial guess atmosphere is carried on for the next cycle. The resulting smoothed low-node inversions were then used as starting guesses for inversions of other line scans close in time. The faculae FAL-C model \citep{1993ApJ...406..319F} was used as the initial atmosphere for every pixel every 5 scans. For the final run, for each scan, the atmospheres were perturbed with the following number of nodes per variable:  Temperature:7, Velocity:3, Microturbulence:1,  By:1, Bx:1, and  Bz:3 (line of sight component). Even though one node was included in Bx and By the main impact of this inclusion for the umbra region was to constrain NICOLE not to generate solutions which would lead to strong Q and U profiles as the signal in Q and U was very low. The weight of the Q and U spectra for the $\chi^2$ computation was half that of Stokes I and NICOLE generally produced atmospheres with low transverse magnetic-field components. Finally, the Stokes V  profiles were given 25\% lower weight than Stokes I for every wavelength. 

Standard inversions in NICOLE usually include a penalty in the $\chi^2$ computation for atmospheres that are not vertically smooth. This is typically referred to as the "regularization". In this work, the final set of analyzed inversions had effectively a regularization of zero. This choice was made to capture vertical variations to the maximum extent allowed by observations in this line (which we believe should be close to that describable by the 7 nodes used for our observations). The disadvantage of such an approach is that seeing induced fluctuations in the line profile, or other noise sources, can potentially be better fitted by a vertical feature and thus bumpier atmospheres can be generated where a smooth atmosphere would sufficiently describe the observed profiles. The latter issue is limited by the low amount of fitting nodes employed. The high quality of the observations, especially with regards to seeing, together with the extended reconstruction scheme and destretch technique of \cite{2012A&A...548A.114H} should also have minimized such effects especially considering that the umbral photospheric structure, used as reference, is well imaged. Furthermore, the combination of two scans at the reconstruction level (as described above) should have led to an unprecedented reduction in seeing signal. The final analysis adds further robustness due to the focus on density plots of a large sample of inverted atmospheres and the usage of two-dimensional maps which are averaged over 1~dex thickness in height (i.e. over a slab corresponding to a difference of 1 in the logarithmic optical depth scale). 

\begin{figure}[!htb]
\begin{center}
\includegraphics{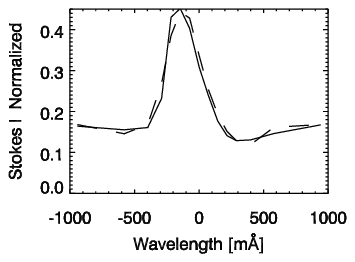}
\includegraphics{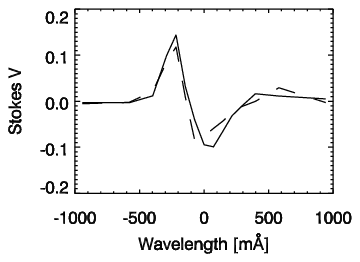}
\includegraphics{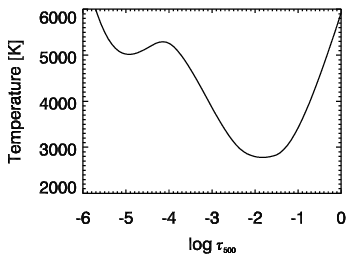}
\includegraphics{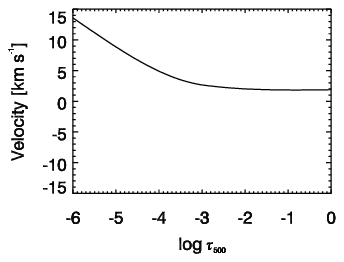}
\end{center}
\caption{Observed (dashed) and synthetized (solid) profiles and atmosphere for one of the most extreme flashes, here reproduced with a hot, strongly downflowing, upper chromosphere. Note the strength of the flash (4x enhancement), the quality of the fit and that, for the Stokes I profile, the minimum is close to 0.10 of the average continuum quiet-Sun intensity.}
\label{detailedmodels}
\end{figure}

\begin{figure}[!htb]
\begin{center}
\includegraphics{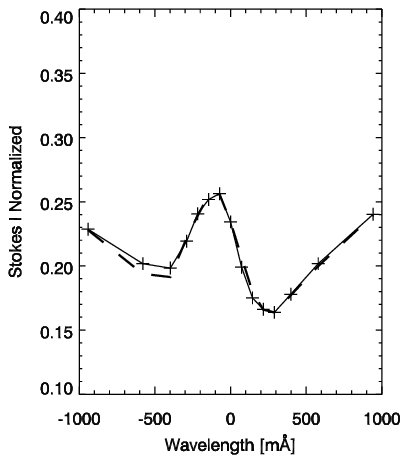} 
\includegraphics{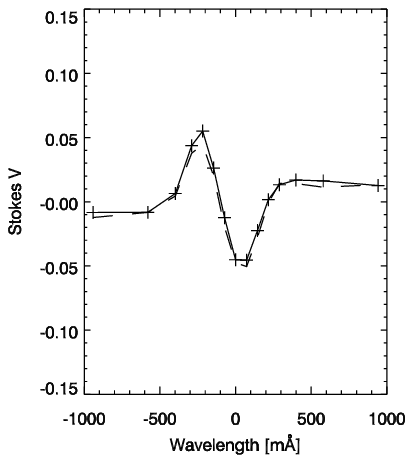}  
\includegraphics{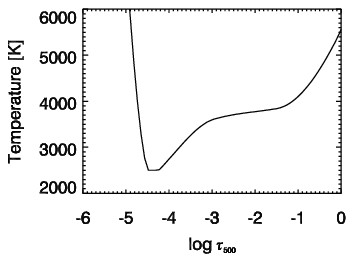} 
\includegraphics{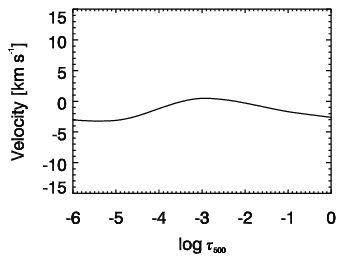}  
\includegraphics{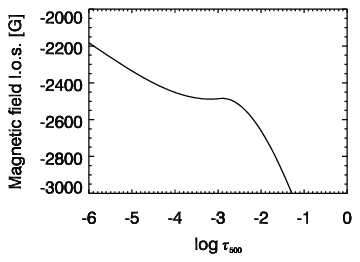} 
\end{center}
\caption{Top row: observed (dashed) and synthetic (solid) profiles for a flashed pixel that was fitted with an upflowing result. Bottom panels: atmospheric properties that generate the synthetic profiles, atmospheric parameters as labeled.}
\label{traditional}
\end{figure}

Since the $\chi^2$ values depend on all wavelengths according to the respective weights in each profile, as well as the regularization, a numerical value for the $\chi^2$ that constitutes a good-enough fit, for the flashed atmospheres, was determined visually from inspecting an array of fits. Any fit with a worse $\chi^2$ than this reference value was selected out from the analysis. Figure~\ref{limit_ths} shows a fit with a $\chi^2$ at the limit of what was considered sufficiently good. The fits are often as good as those shown in Figures \ref{detailedmodels}, \ref{flashdouble}, and \ref{traditional}, discussed later in this work. 

A portion of the upper side umbra and left side penumbra were inverted. In this paper we focus on the umbra, which was selected via intensity thresholding in the line wings followed by an erode morphological operation, performed in order to reduce the area of the mask and minimize the impact of straylight from the penumbra (see Figure~\ref{context}). A total of 37 scans were inverted and analyzed, each 28 seconds long. The inverted pixels were labelled as flashed pixels if the maximum intensity of the flash spectral feature (i.e. the intensity at any point between -217 mA and -73 mA) was higher than that of the far red wing (+942~m\AA). Figure~\ref{context} shows the locations of the atmospheres selected in this way. After selecting for quality of fit, the population of flashed column atmospheres numbers 878. A total of 10989 pixels were considered as quiescent atmospheres by selecting spectra where the maximum intensity between -217~m\AA\ and -73~m\AA\ was under 0.12 of the fitted continuum intensity across random evenly sampled range of scans. The temperature and velocity profiles, as a function of height, were composed into density plots as shown in Figure~\ref{densityplots}. Other atmospheres from the literature are over-plotted for comparison and reference during the discussion. Similarly, magnetic profiles are composed in the density plots of Figure \ref{bdens}.

The dark fibrils in the flashed areas were identified and studied in two different ways: one by manually searching the data cube with CRISPEX \citep{2012ApJ...750...22V} for dark streaks in the flashes and then examining the corresponding atmospheres in the inversion cubes. The atmosphere and profiles shown in Figure~\ref{ucf} are taken from this analysis. A second identification was performed by taking flash masks and closing any gaps in them with a morphological close operation (with a 3 pixel radius circular kernel or 0"35 radius), and then simply subtracting the original flash mask. The flash masks used for this purpose were produced using an intensity criteria lower than that used for flash analysis, namely a pixel is considered to be flashed if any profile point between -217 mA and -73 mA is brighter than the profile at +398m\AA. This procedure, together with the small morphological mask, objectively detects small dark fibrils enveloped by flashes. In Figure~\ref{context2} these are visible as the counterparts of the dark holes present in the flashed mask of the same figure. The density plots for the dark fibrils (see Figure~\ref{ucf} and Fig~.\ref{bdens}) were performed on the population so obtained. These comprise of 448 dark fibril atmospheres.

\section{Analysis}
\subsection{Radiative transfer results}
\label{radtransfer}

\begin{figure}[!htb]
\begin{center}
\includegraphics{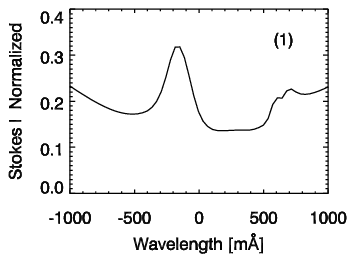} 
\includegraphics{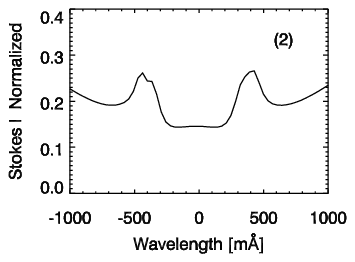}  
\includegraphics{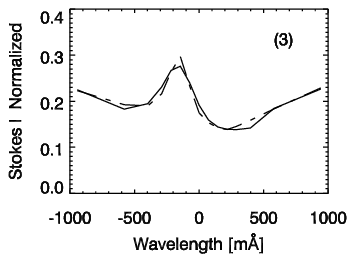} 
\includegraphics{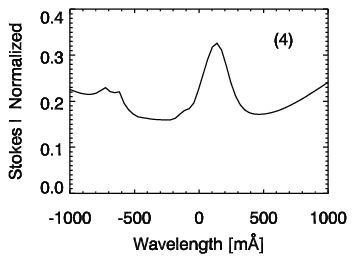}
\end{center}
\caption{Top left: a synthesized Stokes I profile from the downflowing atmosphere shown in Figure~\ref{detailedmodels}. This profile has no instrumental or sampling effects applied. Top right: synthesis in the same atmosphere but with all velocities zeroed. Bottom left: same as top right but with instrumental effects applied, dot-dashed is the observed profile. Bottom right: same as top left but with the sign of the velocities swapped.}
\label{radtransf}
\end{figure}

Some of the flash profiles with a blueshifted emission core are fitted by the
inversion code with a model that has strong downflows in the
chromosphere, especially close to the transition region, and a hotter temperature profile with a lower gradient than that of the transition region. The fits are good, indicating that the downflowing
model is a proper solution to the inverse problem. However, this would
conflict with previous works, both theoretical and observational,
which invariably conclude that flashes are the result of upflowing material resulting from shocking upward-propagating waves.
Moreover, from the point of view of radiative transfer, it is very
difficult to understand how a downflowing chromosphere could produce a
blueshifted feature.

Some research into this issue revealed an interesting synthesis result. Figure~\ref{radtransf} (panel 1) shows the synthetic profile
produced by NICOLE with a blueshifted emission from a downflowing
model. This profile has been computed with the fine wavelength
sampling used internally in the computation. The observations have a
coarser sampling and, with the instrumental profile, we obtain the profile shown in panel 3 which reproduces very well the observed profile of the flash shown in Figure~\ref{detailedmodels}.

 If we now take the model and artificially switch off
the velocity, we obtain the profile in panel 2. Notice that, instead
of a single emission feature, this model is producing 
emission with a central self-absorption, similar to what is often
observed in \ion{Ca}{2}~H and K, but with a deep and flat
absorption core. In this situation we actually have two peaks. Now,
with a suitable velocity gradient, it is possible to introduce a peak
asymmetry such that the red peak is almost completely washed out and
only the blue peak remains visible. Notice that the red peak was still
visible in the first synthetic profile (panel 1) but, because of its
lower amplitude, not only is it weaker but it is also almost
completely lost when convolved with the instrumental profile and resampled to match the observations (panel 3). If we artificially flip the sign of the velocity, we
obtain the opposite effect, with the blue peak disappearing and leaving the red one (panel 4). This is an interesting radiative transfer result in itself as one can have atmospheres generating apparent single emission profiles with Doppler shifts of opposite signal to that actually present. Although not in flashes, emission features with Doppler shifts opposite to the actual flow have been found before \citep{1984mrt..book..173S,2015ApJ...810..145D}. In those works it was found that strong flows were shifting opacity from the red wing into the blue wing, thus causing emission features in the red to be enhanced and similar features in the blue to dampen (and also vice-versa in \cite{1984mrt..book..173S}). A strong velocity gradient, starting at the upper layers, was a key ingredient and the same effect was found to play a role in the peak asymmetries seen in Ca~II grains and flares \citep{1997ApJ...481..500C,2015ApJ...813..125K}. 

With higher spectral sampling and for this particular observed profile, perhaps we would have found that such a small red peak was not present, thus having enough constraint in the inversions so that the downflowing solution would not be selected. On the other hand, this is a particularly extreme pixel in terms of flash intensity, with an enhancement of 4 times the quiescent intensity (normal values being around 2). It may be that, if such hot downflows are present in the umbra of sunspots, the red attenuated peak will be very difficult to detect with any instrumental setup. Further, note that, for this particular example, we do not have an upflowing atmosphere that reproduces the observed profiles equally well even if such solution exists.

If one decides, based on previous literature, that the downflowing solutions are not real, then one can penalize those solutions in the $\chi^2$ computation or outright remove them from the analysis. Perhaps a better way to resolve ambiguities regarding atmospheres that are very different in the upper chromosphere or lower transition region (see full discussion of their properties in Section~\ref{analysis:atmos}), may be to simply use transition region or upper chromospheric diagnostics, such as the ones available with IRIS, by inputing additional temperature or velocity constraints directly into the chromospheric inversions, or fully inverting them together. Given that the downflowing atmosphere is about 500~K hotter than the upflowing solution, just below the transition region, it may be possible to decide between the two families of models with an additional temperature diagnostic that samples such heights, without having to invert multiple lines together.  

\begin{figure}[!htb]
\begin{center}
\includegraphics{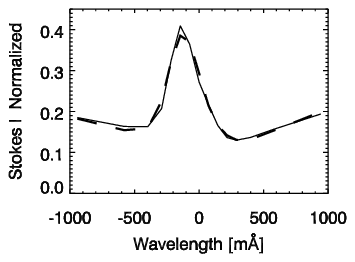}
\includegraphics{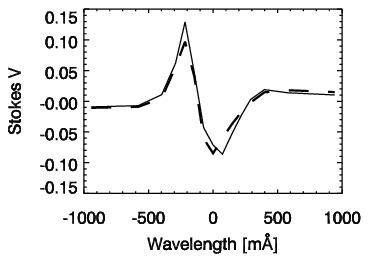}
\includegraphics{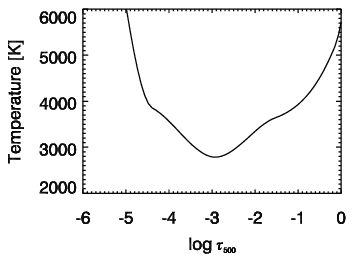}
\includegraphics{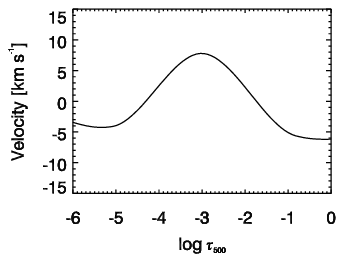}
\end{center}
\caption{Top row: observed Stokes I and V profiles (dashed) and respective fits (solid) for a flashed pixel. The bottom panels show the fitted atmosphere, featuring a velocity stratification with both up and downflows. }
\label{flashdouble}
\end{figure}

Finally, we have obtained a third set of solutions. In low node inversions these look similar to the downflowing family of solutions, but when increasing the number of nodes to 7, the fits become considerably better with a second inflection from a downflow to an upflow above $\log \tau_{500} = -4$. The peak of the downflows in such atmospheres tend to be around where quiescent and flashed atmospheres diverge, at $\log \tau_{500} = -3$. An extreme example of such a fit and respective atmosphere is shown in Figure~\ref{flashdouble}. Due to the reduced sensitivity of the line in the upper layers, this solution may not be indicating any real flows aside from the purely downflowing solution. The presence and proper modeling of NLTE effects does provide some sensitivity to layers all the way up to the transition region as different incoming radiation fields from above will lead to different population levels. However, this sensitivity is mostly in temperature.

In this paper we discuss both main families of solutions (downflowing and upflowing) separately. 

\subsection{Inversion results of flashed profiles}

\begin{figure}[!htb]
\begin{center}
\includegraphics{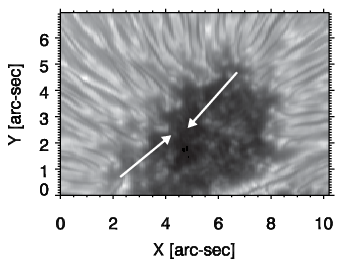}
\includegraphics{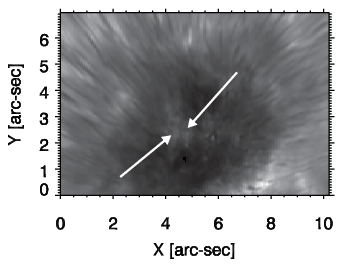} 
\includegraphics{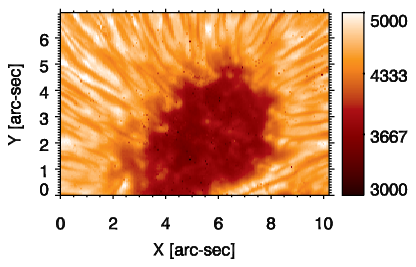} 
\includegraphics{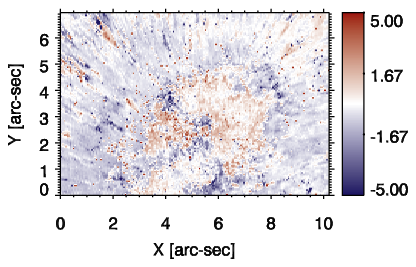} 
\includegraphics{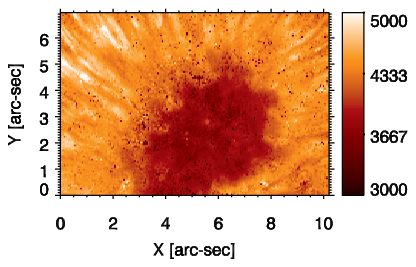} 
\includegraphics{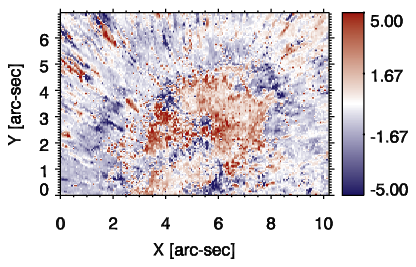} 
\includegraphics{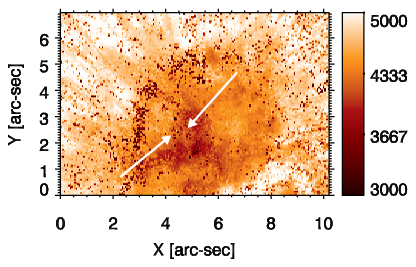} 
\includegraphics{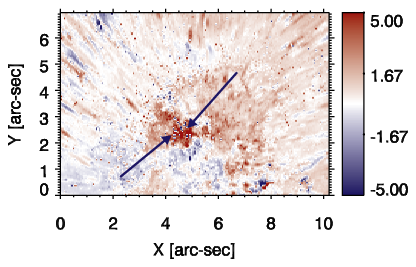} 
\includegraphics{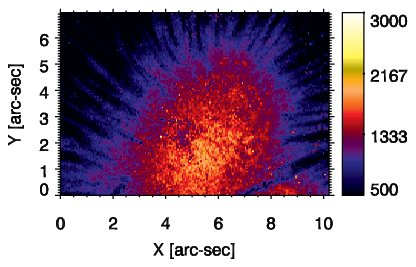} 
\includegraphics{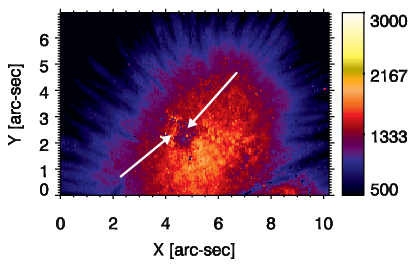}
\end{center}
\caption{Top row : same images as Figure~\ref{context} but with arrows added. Left column: inverted temperature maps. Right colum: respective velocity maps. Second row: the mean over a slab from the upper photosphere to lower chromosphere ($\log \tau_{500} = -1.5$ to $-2.5$).  Third row: the mean taken over $\log \tau_{500} = -2.5$ to $-3.5$.  Fourth row: mean from $\log \tau_{500} = -3.8$ to $-4.8$. Bottom row left: magnetic field averaged over $\log \tau_{500} = -1.5$ to $-2.5$. Bottom row right: magnetic field map averaged from $\log \tau_{500} = -3.8$ to $-4.8$. The left arrow indicates an area of enhanced magnetic field in the upper layers. The right arrow indicates an area of reduced magnetic field.}
\label{tmaps}
\end{figure}

In Figure~\ref{tmaps}, we show temperature and velocity maps for three different layers for a scan where multiple flash fronts are progressing. Diffuse regions with velocities around 1~km~$s^{-1}$ for the upper photospheric layer ($\log \tau_{500} = -1.5$ to $-2.5$) are visible. The contrast is generally low at such heights.  As in \cite{2013A&A...556A.115D}, the inner penumbra is mostly upflowing. As one goes up to $\log \tau_{500} = -2.5$ to $-3.5$ in height, the amplitude of all flows seems to increase and so does the contrast. At this height the umbra shows a mixed picture of up and downflows. Progressing higher to between -3.5 and -4.5, the inner penumbra shows a higher abundance of downflows, in the 1~km~$s^{-1}$  range, and the whole map has, again, lower contrast. Some relation between strong downflows and hotter patches is visible in these maps. 

\subsubsection{Downflowing atmospheres}
\label{analysis:atmos}

As shown in the first row of density plots of Figure~\ref{densityplots}, where darker indicates more points, the inverted atmospheres split into two clear branches in the inverted velocity. In the second row we selected all atmospheres with downflows stronger than 3~km~$s^{-1}$ (a total of 527 atmospheres). The downflowing flashed solutions feature a hot region that slopes up from $\log \tau_{500} = -3$ to $\log \tau_{500} \sim -4.5$, with a much lower gradient than that of the transition region. The relative increase compared with the upflowing flash solutions and the average quiescent umbra atmospheres is 1000~K at $\log \tau_{500} = -4$. This is about the same relative increase as in \cite{2013A&A...556A.115D} even if, in that work, the effect of the flash in temperature was more of a flattening with no atmospheres going above 4000 K before the transition region (similar to the upflowing results in this paper). In both cases the atmospheres diverge at $\log \tau_{500} = -3$. The velocity stratification slopes from just over a 10~km~$s^{-1}$ downflow at $\log \tau_{500} = -5$ to close to rest at $\log \tau_{500} = -3$. 

\begin{figure}[!htb]
\begin{center}
\includegraphics{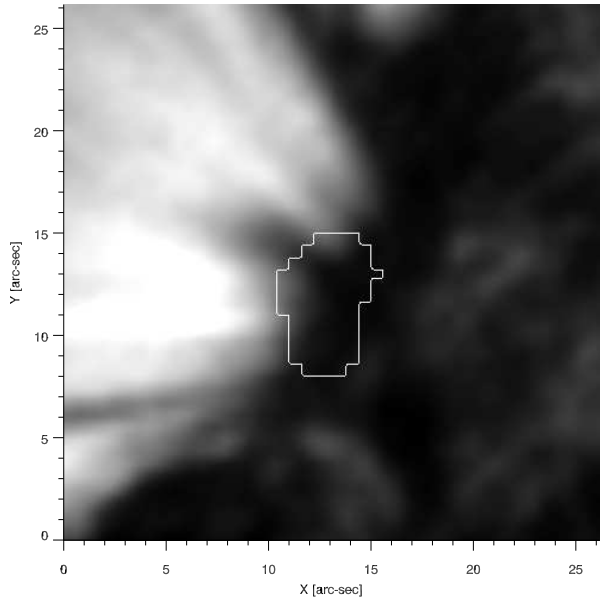}
\end{center}
\caption{Plume structure visible in AIA 171. The bright contour outlines the dark and largest umbra as traced from AIA's 1700 passband (simple intensity masking) and the upper two thirds match approximately with the umbra contours shown in Figure~\ref{context}.}
\label{aia}
\end{figure}

This is the first report, to our knowledge, of a strongly downflowing semi-empirical atmosphere, in the upper chromosphere, for the umbral flashes. Looking at other aspects of this sunspot one finds, rooted in the umbra, the footpoint of a coronal loop (or footpoints of multiple loops) as observed in the EUV in the 171~\AA\ bandpass with AIA (see Figure~\ref{aia}). Such loops are known to be strongly downflowing in the lower corona and transition region, and are most likely what earlier literature calls sunspot plumes \citep{2008AnGeo..26.2955D}, regions of enhanced emission in lines formed in the $10^5$~K to $10^6$~K range and typically colder than the surrounding corona (e.g. \cite{1985ApJ...297..805N,2005ApJ...622.1216B}). Such plumes occur in a majority of sunspots (but not in all and rarely rooted in the umbra) with downflows of up to 25~km~s$^{-1}$ \citep{1999SoPh..190..437M}. The relation between downflows in plumes and higher empirical temperatures seems to be causal as both parameters tend to co-evolve in time \citep{2005ApJ...622.1216B}.

Downflows directly over the umbra in the transition region have been observed as early as \cite{1982SoPh...77...77D} with 5 to 20~km~$s^{-1}$ steady flows and up to 150~km~$s^{-1}$ in localized channels. The early review of \cite{1997ASPC..118...91M} shows that such observations were not an isolated case. More recently, extreme downflows up to 200~km~$s^{-1}$ have been observed in the transition region above sunspots by \cite{2014ApJ...789L..42K} which were interpreted as coronal rain. H-alpha observations of coronal rain above sunspots have been shown to have average velocities of 60~km~$s^{-1}$ \citep{2012ApJ...750...22V}. A recent strong downflow event in the umbra of a spot, displaying EUV emission with properties consistent with those of plumes, has been detected in both transition region and chromospheric lines by \cite{2016ApJ...821L..30K} with evidence for excitation of the chromospheric 3-minute oscillations. Furthermore, on a spot displaying a coronal loop rooted in the dark umbra, similar to the one in this work, \cite{2016A&A...587A..20C} found supersonic downflows in the transition region and in the upper-chromospheric line of Mg~II~k~2796~\AA\ of around 100~km~$s^{-1}$. Perhaps more interesting is that they found a velocity of 15~km~$s^{-1}$ in the transition region lines at the very footpoint of the structure. They found such value to be consistent with a post-shock flow, in terms of mass flux conservation and following from the Rankine-Hugoniot condition for isothermal shocks, from the 100~km~$s^{-1}$ flow. Such downflow of 15 km~$s^{-1}$ is just above our top-most  inverted values for the transition region in the downflowing family of solutions, which can be seen in Figure~\ref{densityplots} at $\log \tau_{500} = -5.5$ (corresponding to the steep transition temperature slope). Our sensitivity at such heights is limited but present (as discussed in Section~\ref{radtransfer}) and the velocity slope is gradual and has very little spread in the density plots. Going further back, evidence for downflows and upflows co-existing within a 0.3~arcsec element (SUMER slit's width) in the transition region of the umbra has existed since \cite{2001ApJ...552L..77B}. If these observations correspond to one component of a siphon flow \citep{1980SoPh...65..251C}, one should remember that, while the inverse Evershed flow is observed well outside the umbra, it would geometrically complement any umbral downflow observed in the higher layers and it has been observed to have amplitudes topping, similarly, 15~km~$s^{-1}$. 

As counter evidence to the idea of strong chromospheric downflows, even in the presence of transition region flows, recently \cite{2015A&A...582A.116S} found a steady supersonic downflow in the transition region of the umbra of a sunspot using IRIS but no chromospheric downflow in the cooler passbands. They find their results compatible with the model of a siphon flow, mainly due to the stability of the observed flows.

Given the evidence from previous works for umbral downflows that extend to the chromosphere and lower transition region, one cannot discard the inverted downflowing solutions as merely a radiative transfer curiosity and has to consider the possibility that the hot-downflow fits are capturing an actual hot, up-ward propagating shock occurring against a strong downflow. It can even be that a second shock from the in-falling material into a higher density layer, similarly to that reported by \cite{2016A&A...587A..20C}, is occurring at the same heights as the flash, and thus interfering or modulating the shocks that would normally occur from the steepening of the 3-minute oscillations. If hot chromospheric downflows are real, it may be that the presence of a sunspot plume, rooted in the umbra, is either critical or greatly increases the possibility of detection. A lot of our confidence in upflowing models comes from synthesis, in hot upflowing atmospheres (upflowing around $\log \tau_{500} = -4$), that successfully reproduce the saw-tooth pattern \citep{2010ApJ...722..888B}. We suggest that a similar procedure should be attempted in future work on the hot downflowing model atmospheres. 

\subsubsection{Upflowing atmospheres}
\label{analysis:atmos}

\begin{figure}[!htb]
\begin{center}

\includegraphics{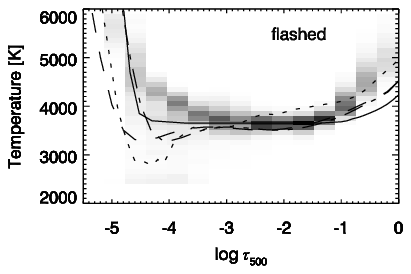} 
\includegraphics{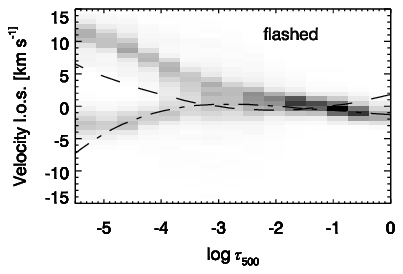} 

\includegraphics{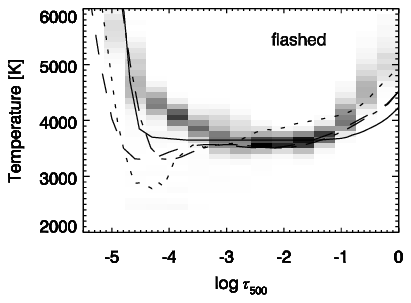} 
\includegraphics{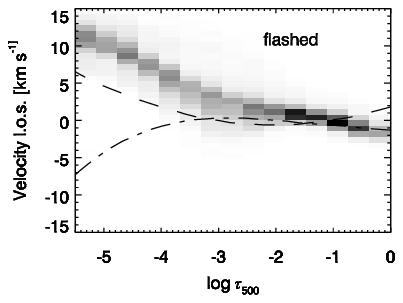} 

\includegraphics{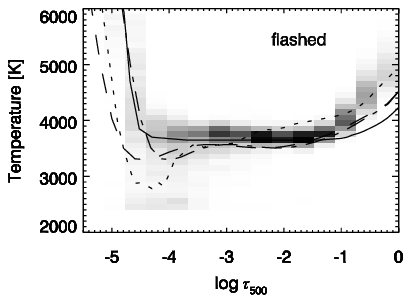}
\includegraphics{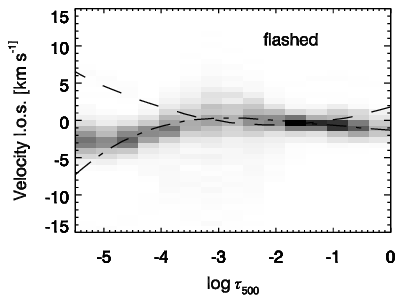}

\includegraphics{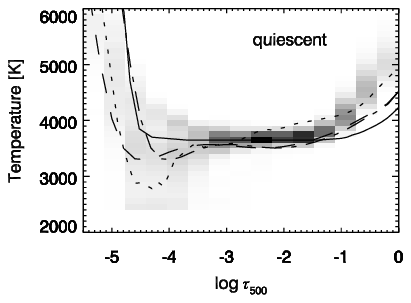} 
\includegraphics{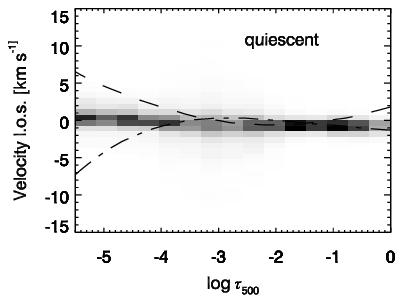}

\end{center}
\caption{Density plots (i.e. darker meaning a higher concentration of points). Left column: temperature stratification versus $\log \tau_{500}$. Right column: velocity stratification versus $\log \tau_{500}$. Top row: flashed atmospheres, all solution types. Second row: flash atmospheres, downflowing solutions only. Third row: flashed atmospheres after downflowing results were filtered out. Fourth row: atmospheric models for the quiescent umbra. Models found in the literature for the same structures and in optical depth scale are over-plotted: solid is MaltbyL, the dotted line is the dark umbra model of  \cite{2007ApJS..169..439S}, the dashed line is the first component of the time-dependent strong-flash model from \cite{2001ApJ...550.1102S} and dash-dot the second component. The dashed velocity profiles are also from the latter two-component model.}
\label{densityplots}
\end{figure}

The upflowing solutions are best seen in the third row of the density plots in Figure~\ref{densityplots} where all solutions with any downflow over 3~km~$s^{-1}$ were removed (for a total of 351 atmospheres). From $\log \tau_{500} = -3.5$ to $-4.5$ these show a spread in temperature that goes from about 4100~K, following a flat top just above the one-dimensional umbral core model ``L'' by \cite{1986ApJ...306..284M} (their table 9; from now on "MaltbyL"), shown as a solid line, to as low as the allowed minimum at 2500~K. The highest density of atmospheres (which the mode of the distribution traces) is close to the upper range, around 3900~K, and seems to match that obtained by \cite{2013A&A...556A.115D} for the same heights. Compared with that work, our spread in temperatures is somewhat higher, but the quantity of the inverted pixels and scans in this work is also much higher. At this height interval, the upflowing flashed atmospheres are hotter than the quiescent case but only considering the mean and the mode. At $\log \tau_{500} = -5$ the solutions are always hotter than any quiescent atmosphere. This appears to be from a shift of the very high temperature-gradient region, corresponding to the beginning of the transition region, to a lower optical opacity, i.e. the transition region is about half a dex deeper in the flashed models than in the quiescent models. This aspect, together with the slight dip in temperature around $\log \tau_{500} = -4$, seems to match the two components from \cite{2000ApJ...544.1141S}, over-plotted in Figure~\ref{densityplots}. The difference is that our flashed and quiescent atmospheres, at the $\log \tau_{500} = -4$ dip, have the highest density of atmospheres at temperatures higher than their respective two-component analogs. Furthermore, our quiescent atmosphere is not strongly downflowing, as is the case with the dashed component from \cite{2000ApJ...544.1141S}. 

A slight enhancement, but only up to a couple hundred K, when compared with both the quiescent case and with the lower heights, is visible around $\log \tau_{500} = -3$. This is where the flashed atmospheres clearly depart from the quiescent case in all variables and for all families of models. It is the same height of divergence as that of  \cite{2013A&A...556A.115D}. This enhancement is very similar to that plotted for all models present in \cite{2000ApJ...544.1141S} but is different in nature to that of \cite{2013A&A...556A.115D} where the divergence between flashed and quiescent atmospheres starts as a change of slope in temperature. For this work, and for both components of \cite{2000ApJ...544.1141S}, there is a visible "bump".  

As far as the velocity is concerned, the value of -5~km~$s^{-1}$ at $\log \tau_{500} = -5$ and the shape of the chromospheric profiles match those obtained by \cite{2013A&A...556A.115D} and \cite{2000Sci...288.1398S}. In velocity it is more visible that the divergence between the families of models starts as low as $\log \tau_{500} = -2$. At the photosphere, even though our sensitivity is limited at such layers, all flashed atmospheres are close to rest. 

\subsection{Magnetic field response}
\label{analysis:atmos}

\begin{figure}[!htb]
\begin{center}
\includegraphics{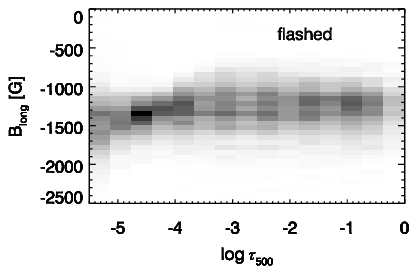}
\includegraphics{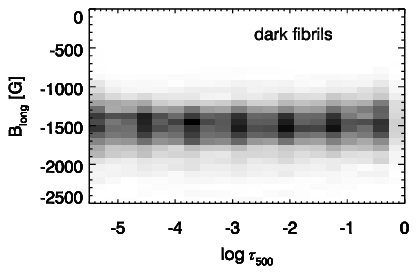}
\includegraphics{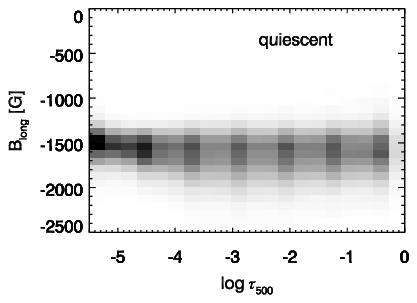}
\end{center}
\caption{Density plots for the magnetic field stratification (line-of-sight magnetic field component versus height in $\log \tau_{500}$). Top left: flashed models. Top right: dark fibril models. Bottom: quiescent models.}
\label{bdens}
\end{figure}

Both in the density plots and the magnetic field maps (see Figs.~\ref{bdens} and the region highlighted by the right arrow in \ref{tmaps}), all families of flashed solutions lead to a reduction in the magnetic field when compared to the quiescent case. The spread of atmospheres prevents a conclusion about stratification itself but an average reduction is clearly present for nearly all flashed pixels when averaging over the column. This leads us to a tentative explanation for the observed magnetic field reduction as an increase in the adiabatic gas pressure from the shock, pushing the magnetic field lines away from the flashed areas.

For some maps one is tempted to infer that there is also a counterpart magnetic field enhancement, at the border of the reduction. One gets the strongest impression at the borders of flashes, such as the one highlighted by the left arrow in Figure~\ref{tmaps}. However, we are unable to claim that this counterpart enhancement is above the spatial inversion noise. 

In a broader context, this work adds to the recent body of evidence for the existence of magnetic field oscillations in sunspots, at least in the chromosphere, \citep{2013A&A...556A.115D}, a debated topic that remains unsolved (e.g. \cite{2010arXiv1012.1196R}). An alternative explanation for the magnetic field reduction is simply observational, in that the normally reduced field in the higher layers (where we could expect some fanning out of the field lines, even if known to be limited in the umbra) is just not captured by the photosphericaly dominated Stokes V profiles unless there is a flash providing some signal from the upper layers. This effect would be similar to that proposed by \cite{2002AN....323..254C,2003ApJ...588..606K} where opacity variations explain much of the small apparent field modulation observed in earlier works. 

\subsection{Inversion results of quiescent profiles}
\label{analysis:atmos}

For the quiescent models (see Figure~\ref{densityplots}), we find that the temperature profiles tend to be flat from $\log \tau_{500} = -1.5$ up to $\log \tau_{500} = -3.5$. Higher up we get a large scatter of models. MaltbyL, plotted as a solid line in Figure~\ref{densityplots}, happens to constitute a good average profile to our scatter, with the exception of MaltbyL being about 100~K cooler between $\log \tau_{500} = -1$ and $-3$. This good match is the reason why we chose to plot this model from the earlier LTE model literature. However, MaltbyL is an unusually hot model when it comes to umbral models for the layers above $\log \tau_{500} = -2$. It is also a relatively flat model (e.g. compare its profile with the other reference atmospheres plotted in Figure~\ref{densityplots}). The chromospheric dark-umbra model of  \cite{2007ApJS..169..439S} (the dotted atmosphere in Figure~\ref{densityplots}) has a more complex stratification, with a dip below 3000~K above $\log \tau_{500} = -4$ which is on the cooler limit of our results. Such a temperature drop below 3000~K in the umbra is supported by the multi-line model of \cite{2009ApJ...707..482F} and the millimeter observations of \cite{2014A&A...561A.133L}, with the caveat that the latter may be valid only for some spots as the brightness in the radio range, similarly to what is observed in the optical, can vary from spot to spot \citep{2016ApJ...816...91I}.  

Similarly to this work, the inversions of high resolution Ca~II~8542 data by \cite{2013A&A...556A.115D} resulted in a wide range of atmospheres for the quiescent phase, with models as cold as 2500~K and a near continuous progression up to about 3500~K at $\log \tau_{500} = -4$. In this work, the quiescent temperature profiles most similar to such models, and to that of \cite{2007ApJS..169..439S}, tend to occur close to the umbra/penumbra boundary, especially in the disk-center side umbra (see Figure~\ref{tmaps}). However, we obtain a clear range of temperatures that peak higher, at just about 4000~K at $\log \tau_{500} = -4$ which was not observed in that previous report. Target-wise, the main observational differences between \cite{2013A&A...556A.115D} and this work are the size of the sunspots, larger in the present work, and the viewing angle ($\mu=0.87$ versus $\mu=1$ in this work). Considering the impact of selection effects, the most unique characteristic of this work is the sheer amount of inverted pixels, with about 10989 quiescent pixels analyzed, which could lead to a larger scatter of atmospheres in the upper layers, but the obtained velocities do not show such scatter. In fact, the scatter of models in velocity seems to be lower than that of \cite{2013A&A...556A.115D}. Finally, the spots were also observed at different phases of the cycle, with the earlier work being close to the minimum of the previous sunspot cycle and this work being just after the maximum. There is evidence that umbral temperatures, at least at photospheric levels, fluctuate with solar cycle \citep[see for e.g.][]{1978Natur.274...41A,1986ApJ...306..284M,2012A&A...541A..60R}. In this sense, due to the similarity of approaches and data, \cite{2013A&A...556A.115D} together with this work add to the body of evidence that the temperature of the umbra of sunspots does vary with cycle, with later sunspots being hotter in the upper chromosphere. This is further reinforced if one considers that the model that best matched our results from the old LTE literature was the late cycle model of \cite{1986ApJ...306..284M}.

In terms of velocity the quiescent phase shows the top end of the distribution downflowing at about 1~km~$s^{-1}$ in the upper layers, but consistently at rest in the photosphere with a very small scatter. The presence of a weak downflow in the upper layers is consistent with the past literature and the idea of a transient strong upflow followed by a slow downflow first put forward by \cite{1969SoPh....7..351B} but, as shown in Figure~\ref{densityplots}, the small scatter of solutions includes the at rest case. 

Magnetic field-wise, as shown in Figure~\ref{bdens}, quiescent atmospheres show a progressive reduction of average field strength and of the range of measured values with increasing heights. The lower end of the distribution reduces in field strength from approximately 2000~G in the photosphere to a narrow range of inverted values  at 1500~G in the upper chromosphere.

\subsection{Inversion results of dark fibrils}
\label{analysis:dark fibrils}

\begin{figure}[!htb]
\begin{center}
\includegraphics{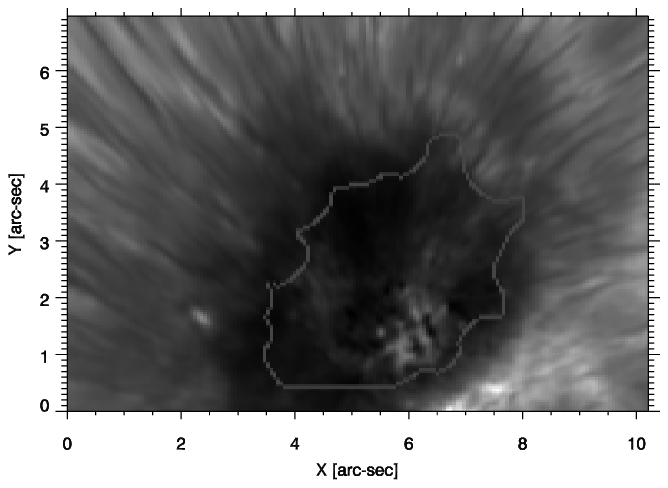}
\includegraphics{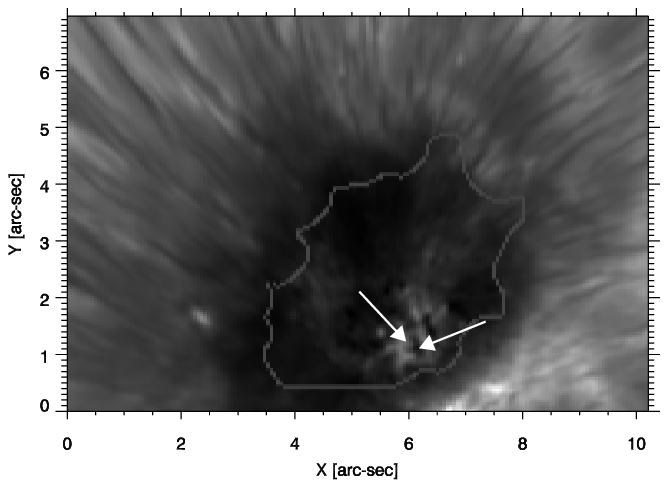} 
\includegraphics{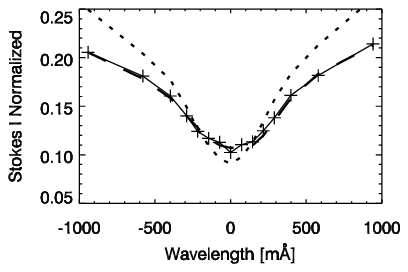}
\includegraphics{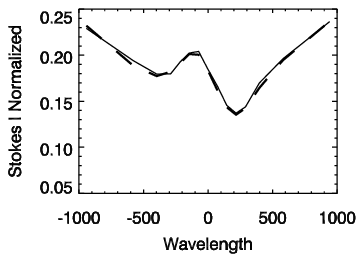}
\includegraphics{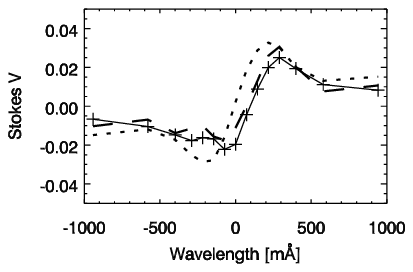}
\includegraphics{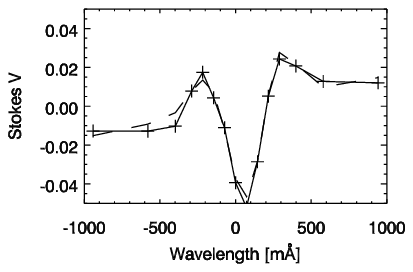}
\includegraphics{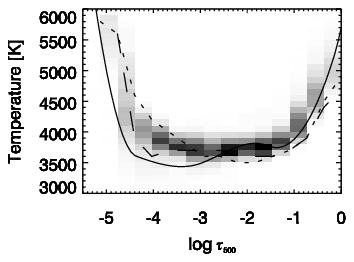}
\includegraphics{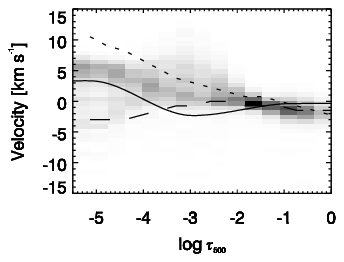}
\end{center}
\caption{Top: the same image of a flash with dark fibrils, with and without arrows, at -73 mA from core. The contour traces the umbra mask. The left arrow points at a pixel in the flashed area and the right arrow points at a pixel in the dark fibril. The Stokes I and Stokes V profiles on the left are for the dark fibril pixel. Those on the right are for the flashed pixel. Dashed lines trace the observed profiles, solid the synthetic profiles. The dotted line is from a typical quiescent atmosphere. The crosses indicate the observed wavelength positions. The bottom row plots the inverted temperature and line of sight velocity for the dark fibril (solid line), density plots for the dark fibrils that were automatically detected (shades proportional to density), the mode of the downflowing flashed atmospheres (short dashes), and the mode of the upflowing flashed atmospheres (long dashes).}
\label{ucf}
\end{figure}

From both the manual and automated analysis, one finds that the typical dark fibril Stokes I profile is not only significantly darker than a flashed profile at the flash-peak wavelengths but also slightly darker in the wings (up to, at least, $\pm~1$~\AA) indicating a photospheric connection. For an example see the profiles in Figure~\ref{ucf}. The profiles of the darkest fibrils are more similar to quiescent profiles than to those of the flashed atmospheres with the Stokes I profiles not showing the flash emission in the line wings, and the Stokes V profiles not showing the abnormal, reversed polarity, blue peak. However, the Stokes V profiles show a flatter blue half and an attenuated red peak when compared to the quiescent case (see Figure~\ref{ucf}), as one would expect if the blue trough had been red-shifted. From the density plots of the inversions (bottom panels of Figure~\ref{ucf}) we find that the observed dark fibril profiles are reproduced with atmospheres that are similar to the flashed atmospheres in terms of velocity and temperature stratifications, albeit with lower amplitudes. Dark fibril profiles are reproduced with flows in the upper chromosphere that are about half those found for the flashed profiles, peaking close to 5~km~$s^{-1}$ at $\log \tau_{500} = -5$. The temperatures are also lower than the flashed case, which might help explain their relative darkness, together with any NLTE effects that NICOLE might be capturing (to be addressed in a future publication). They are hotter than the average quiescent atmosphere. The majority of the dark fibrils detected are downflowing. Considering that a good portion of the observed dark fibrils should be short dynamic fibrils, known to have up- and down-flowing phases \citep{2013ApJ...776...56R}, one must be open to the possibility that something about our intensity detection method is preferentially selecting the downflowing stages. It may also be that short dynamic fibrils are the darkest during their downflowing stage when observed at $\mu=1$. 

The inverted average magnetic field strength (see Figure~\ref{bdens}) seems to occupy the values between those obtained for the quiescent and the flashed case. The exception is the magnetic field above $\log \tau_{500} = -4$ where both dark fibrils and flashed atmospheres don't show the same progressive drop with height that the quiescent atmospheres do. 

These results are consistent with dark fibrils being primarily caused by atmospheric inhomogeneities that affect the propagation of the flash. Since we are likely resolving the same two-components put forward by \cite{2000Sci...288.1398S}, it makes sense that at least one of the components here observed would have stronger amplitudes in velocity when compared to the previous literature. As discussed above, upper-layer flows from an unusual loop rooted in the umbra may be partially responsible in generating stronger and more easily identifiable components (at least downflowing ones). Generalizing to other observed sunspots, the existence of a different flow structure across the umbra would provide the inhomogeneity that allows for the existence of short-dynamic fibrils/spikes and dark umbral fibrils in the first place. One scenario is that the upper-layer flow modulates the height at which the shock-front occurs. More specifically, in such a scenario, from the frame of reference of the downflowing material, the shock that generates the umbral flash would still be propagating upwards in an isotropic wave as generally understood, but from an external point of view, a visible shock propagating along such a downflowing region would appear lower than one propagating in the upflowing regions. This would lead to visually distinct features separate in height. Furthermore, if the velocity gradient is sufficiently low (i.e., if a downflow extends deep enough into the chromosphere), then a downflow could hide the flash brightening into a deeper layer, causing the profile to show both a reduced blue emission peak, and a red-shifted line-core, exactly like the one shown in Figure~\ref{ucf}. 

\section{Concluding remarks}%%%%%%%%%%%%%%%%%%%%%%%%%%%%%%%%%%%%%%%%%%%%%%%%%%%%%%%%%%%%%

We present NLTE inversions of chromospheric high-resolution spectro-polarimetric observations of umbral flashes, discriminated by type of feature. We find two families of solutions for umbral flashes. One is upflowing in the chromosphere with the transition region moving lower in height when compared with the quiescent phase, in line with previous results. The second is strongly downflowing, with downflows in excess of 5~km~$s^{-1}$ in the upper chromosphere and increasing up to just under 15~km~$s^{-1}$ in the transition region. These feature a region with lower temperature gradient in the chromosphere but hotter than the upflowing family of solutions at such heights. This new family of solutions for flashes is either a radiative transfer effect that may or may not have counterparts in the Sun (and thus a pitfall to watch out for and invalidate) or a real solution reported here for the first time. Given the presence of a coronal loop, visible in the EUV 171 \AA\ bandpass with AIA (up to $10^5$ K), and the previous literature on the flows of these structures as well as recent discoveries of downflows in the chromosphere of the umbra of other sunspots, we propose that both the downflowing and the upflowing solutions are real and extremes of a commonly occurring inhomogeneity in flows. Such inhomogeneity, at smaller scales, would help explain the visibility of phenomena such as short-dynamic fibril, spikes and dark features observed in flashes in general. Such downflows and their presence in dark fibrils also provide a direct piece of the puzzle in the recently discovered relationship between short dynamic fibrils and umbral microjets \citep{2017arXiv170505617N}.  

We also provide models for the quiescent umbra that are, on average, flatter and hotter in temperature stratification than most previous literature but that agree with the late-cycle umbral model of \cite{1986ApJ...306..284M}. With this work focusing on a late cycle sunspot, together with \cite{2013A&A...556A.115D} obtaining lower average temperatures with a similar study for an early-cycle spot, this constitutes additional evidence for the hypothesis that the umbrae of late-cycle sunspots, at least in the chromosphere, are indeed hotter than those of early-cycle sunspots. This result is also the first chromospheric NLTE empirical study in support of the Maltby late-cycle umbra model. 

We find supporting evidence, from the NLTE inversions, for the previous LTE literature on umbral models that seemed to indicate that the magnetic field in the umbra only has a slight drop in strength from the photosphere to the chromosphere. Perhaps more importantly, we find that flashed areas show lower magnetic field strengths on average. Speculatively, the latter may be due to an increase in gas pressure with the shock pressure pushing field-lines away. Alternatively, higher sensitivity to the upper layers of the atmosphere, caused by the flash, may lead to success in reproducing an ever present reduction of the field with height.

We find that the darkest fibrils observed in the flashes have Stokes profiles that are distinct from both flashed and quiescent regions. While similar to those of the quiescent umbral atmosphere, clear differences are invariably fitted with flash-like thermodynamical properties albeit with reduced amplitudes. Given the ensemble of results presented in this paper, it is tempting to interpret the observed dark fibrils as a manifestation of an inhomogeneous flow structure in the umbra, possibly affected by transition region or even coronal flows, affecting the speed at which the shockwave from the flash propagates from an external point of view. 

As a final note, it is interesting that for any family of solutions, bright flashed atmospheres and their dark features are always strongly flowing, whereas quiescent atmospheres are always very close to rest.

\begin{acknowledgements}
We would like to thank the anonymous referee for relevant questions that led to improvements in the quality of the results. We would also like to thank Peter S\"{u}tterlin for assisting with the observations. We acknowledge support from Robert Ryans with computing infrastructure. This work made use of Nicola Vitas online repository of atmospheres (http://nikolavitas.blogspot.no/2014/10) and Simon Vaughan's density plot routine available at (https://github.com/svdataman/IDL). The Swedish 1-m Solar Telescope is operated on the island of La Palma by the Institute for Solar Physics (ISP) of Stockholm University
in the Spanish Observatorio del Roque de los Muchachos of the Instituto de Astrof\'isica de Canarias. This work was performed using the Darwin Supercomputer of the University of Cambridge High Performance Computing Service (http://www.hpc.cam.ac.uk/), provided by Dell Inc. using Strategic Research Infrastructure Funding from the Higher Education Funding Council for England and funding from the Science and Technology Facilities Council. This research was supported by the SOLARNET project (www.solarnet-east.eu), funded by the European Commissions FP7 Capacities Program under the Grant Agreement 312495. This work has been supported by the UK Science and Technology Facilities Council (STFC) and partially supported by the Norwegian Research Council (project 250810 / F20). JdlCR is supported by grants from the Swedish Research Council (2015-03994), the Swedish National Space Board (128/15) and the Swedish Civil Contigencies Agency (MSB). 
\end{acknowledgements}

\end{document}